\documentclass[conference]{IEEEtran}
\IEEEoverridecommandlockouts
\usepackage{cite}
\usepackage{amsmath,amssymb,amsfonts}
\usepackage{algorithm}
\usepackage{algpseudocode} 
\usepackage{graphicx}
\usepackage{textcomp}
\usepackage{color}
\usepackage{hyperref}
\usepackage{fontawesome}
\usepackage{booktabs}   
\usepackage{siunitx}    
\usepackage{tabularx} 
\usepackage{makecell}   
\usepackage{array}      
\usepackage{multirow}
\usepackage{cite}
\newtheorem{definition}{Definition}
\def\BibTeX{{\rm B\kern-.05em{\sc i\kern-.025em b}\kern-.08em
    T\kern-.1667em\lower.7ex\hbox{E}\kern-.125emX}}
\begin{document}

\title{Pilot-guided Multimodal Semantic Communication for Audio-Visual Event Localization}
\author{Fei Yu~\IEEEmembership{Memember,~IEEE,} Zhe Xiang, Nan Che,~\IEEEmembership{Memember,~IEEE,} Zhuoran Zhang,\\ Yuandi Li, Junxiao Xue, Zhiguo Wan}
\maketitle

\begin{abstract}
Multimodal semantic communication, which integrates various data modalities such as text, images, and audio, significantly enhances communication efficiency and reliability. It has broad application prospects in fields such as artificial intelligence, autonomous driving, and smart homes. However, current research primarily relies on analog channels and assumes constant channel states (perfect CSI), which is inadequate for addressing dynamic physical channels and noise in real-world scenarios. Existing methods often focus on single modality tasks and fail to handle multimodal stream data, such as video and audio, and their corresponding tasks. Furthermore, current semantic encoding and decoding modules mainly transmit single modality features, neglecting the need for multimodal semantic enhancement and recognition tasks.

To address these challenges, this paper proposes a pilot-guided framework for multimodal semantic communication specifically tailored for audio-visual event localization tasks. This framework utilizes digital pilot codes and channel modules to guide the state of analog channels in real-wold scenarios and designs Euler-based multimodal semantic encoding and decoding that consider time-frequency characteristics based on dynamic channel state. This approach effectively handles multimodal stream source data, especially for audio-visual event localization tasks. Extensive numerical experiments demonstrate the robustness of the proposed framework in channel changes and its support for various communication scenarios. The experimental results show that the framework outperforms existing benchmark methods in terms of Signal-to-Noise Ratio (SNR), highlighting its advantage in semantic communication quality.

\end{abstract}

\section{Introduction}
Semantic communication aims to achieve accurate interaction of semantic information between the sender and receiver \cite{bao2011towards}. By leveraging advanced artificial intelligence (AI) techniques, it extracts the most relevant information from raw data for specific intelligent tasks. This approach effectively compresses data redundancy, improves information transmission efficiency, alleviates network transmission pressure, and reduces processing delay for intelligent tasks \cite{strinati20216g,luo2022semantic}.

Multimodal semantic communication, a subset of semantic communication, integrates various data modalities such as text, images, and audio, significantly enhancing communication efficiency and reliability \cite{zhang2024unified,chaccour2024less}. This advancement demonstrates broad application prospects and importance in fields such as artificial intelligence, autonomous driving, and smart homes. By merging different types of data, multimodal semantic communication enriches the conveyed information and strengthens the system's capability to handle complex tasks \cite{wang2023multimodal,luo2022multimodal}. For instance, in the multimodal audio-visual event localization (AVE) tasks~\cite{ave,psp}, multimodal communication enables the integration of visual and auditory data to accurately identify and locate events, which is crucial for applications requiring detailed context and precise localization.

\noindent \textbf{Current Issues.} Despite the broad potential of multimodal semantic communication, current research often relies on analog channels and assumes a constant channel state, which presents limitations when addressing the dynamic challenges of physical channels and noise in real-world scenarios \cite{wang2024distributed,xie2024hybrid}. Existing methods typically focus on single-modality tasks and have not fully addressed the handling of multimodal stream data, such as video and audio, and their corresponding tasks \cite{xie2021task,xie2022task,10226153,liao2024adasem}. Additionally, current semantic encoding and decoding modules primarily transmit single-modality features, failing to consider the needs for multimodal semantic enhancement and recognition tasks. Moreover, existing multimodal semantic communication methods often overlook the potential for inter-modal information complementarity and enhancement to mitigate information loss due to physical noise and semantic encoding/decoding. This limitation affects the ability to handle tasks requiring synchronization and interpretation across multiple data types. Therefore, there is a need for a framework that can adapt to varying channel conditions and support comprehensive integration of multiple data modalities.

\noindent \textbf{Challenges.} In the rapidly evolving field of multimodal data processing, several critical challenges persist. These challenges stem from the limitations of current methods and the complexity of real-world applications:
\begin{itemize}
    \item Handling Multimodal Stream Data: Current multimodal data processing methods often focus on integrating static modalities such as images and text. However, they fall short when dealing with dynamic multimodal stream data, such as video and audio, which require synchronization and real-time interpretation across multiple data types \cite{wang2023multimodal,luo2022multimodal}. This limitation affects the ability to process tasks that involve continuous, synchronized data streams effectively.
\item Semantic Enhancement Across Modalities: Existing research on semantic encoding and decoding primarily addresses single-modality tasks~\cite{xie2021task,xie2024hybrid,xie2022task}, focusing on transmitting features from individual modalities without fully integrating them. This oversight results in a lack of comprehensive multimodal semantic enhancement, which is crucial for tasks requiring nuanced interpretation and interaction between modalities.
\item Dynamic Channel Conditions: Traditional methods~\cite{wang2024distributed,xie2024hybrid} often assume static or constant channel conditions, which is inadequate for real-world scenarios where channel characteristics vary dynamically. This gap impacts the system’s robustness and accuracy in environments with fluctuating signal conditions.
\end{itemize}


\noindent \textbf{Proposed Research Solutions.} To address these challenges, this paper proposes a pilot-guided framework for multimodal semantic communication specifically tailored for audio-visual event localization tasks. This framework utilizes digital pilot codes and channel modules to guide the state of analog channels in real-time and designs Euler-based multimodal semantic encoding and decoding that considers time-frequency characteristics based on real-time channel state. This approach effectively handles multimodal stream source data, particularly for audio-visual event localization tasks. 

\noindent \textbf{Summary of Novel Contributions:}
\begin{itemize}
\item \textbf{Pilot-guided Channel State Estimation}: We propose the first pilot-guided multimodal semantic communication framework, which uses pilot signals to accurately estimate dynamic channel conditions. This approach simplifies system design, enhances robustness, and provides reliable channel information, outperforming traditional methods that assume constant channel states.
\item \textbf{Multimodal Semantic Enhancement}: We propose the cross-terminal multimodal semantic-enhancement encoder/decoder, which integrates and enhances information across multiple modalities. Unlike existing methods, our framework addresses the critical gap in semantic enhancement across single modalities in multi-user settings, effectively mitigating information loss and noise impacts during semantic transmission.
\item \textbf{Euler Time-Frequency Characteristics}: We use Euler's formula to map features from 2D to angular space, which significantly improves accuracy in audio-visual event localization (AVE) tasks compared to existing methods under zero-loss channel transmission. Our approach enhances semantic encoding and decoding, offering notable advantages in accuracy while maintaining robust performance even in noisy communication scenarios.
\end{itemize}
Extensive numerical experiments demonstrate that the framework exhibits good robustness under varying channel conditions. The results show that the method outperforms existing techniques in terms of Signal-to-Noise Ratio (SNR), highlighting its advantages in semantic communication quality.

\section{Related Work}
\subsection{Multimodal Semantic Communication}
Semantic communication aims to transmit semantic information accurately between sender and receiver \cite{bao2011towards}. By leveraging advanced AI techniques, it extracts and communicates the most relevant data, enhancing transmission efficiency, reducing redundancy, and minimizing delays \cite{strinati20216g,luo2022semantic,9797984}.

Multimodal semantic communication extends this concept by integrating multiple data modalities—such as text, images, and audio—to further improve communication efficiency and reliability \cite{zhang2024unified,chaccour2024less}. This integration offers significant benefits across fields like artificial intelligence, autonomous driving, and smart homes, by enriching information exchange and enhancing the system's ability to manage complex tasks \cite{wang2023multimodal,luo2022multimodal,10225839}.

However, existing research often relies on analog channels with constant channel state assumptions, which are insufficient for addressing the dynamic nature of real-world channels and noise \cite{wang2024distributed,xie2024hybrid}. Current methods generally focus on single-modality tasks and do not fully address the challenges of handling multimodal stream data, such as video and audio \cite{xie2021task,xie2022task}. Additionally, existing semantic encoding and decoding modules tend to handle single-modality features and overlook the need for multimodal semantic enhancement and recognition. Many approaches also neglect the benefits of inter-modal information complementarity for mitigating information loss due to physical noise and semantic encoding/decoding issues. This underscores the need for frameworks that can adapt to varying channel conditions and support comprehensive multimodal integration.

Previous work has addressed various aspects of multimodal communication. For instance, Xie et al. \cite{xie2021task} explored a multi-user system for image and text transmission aimed at visual question answering (VQA). Xie et al. \cite{xie2022task} introduced a Transformer-based structure for text and image transmission and proposed several deep learning-based frameworks for tasks such as image retrieval and machine translation. Li et al. \cite{li2022cross} proposed a cross-modal paradigm leveraging complementary information to enhance communication reliability. Luo et al. \cite{luo2022multimodal} developed a multi-modal data fusion scheme specifically for wireless channel characteristics. Despite these advancements, challenges remain, including mitigating noise from wireless channels, managing reception under unknown channel conditions, designing effective transmission models for specific tasks, and enhancing semantic relevance between modalities.

\subsection{Audio-Visual Event Localization}
Tian et al. \cite{tian2018audio} introduced the concept of audio-visual event localization, where an audio-visual event is defined as an event that is both audible and visible within a video clip. Audio-visual event localization is a challenging task that requires establishing effective cross-modal connections between audio and video segments to learn robust representations of both modalities. It also necessitates a carefully designed understanding of video content to accurately predict event categories and discern event segments from distracting backgrounds. Early studies \cite{tian2018audio,lin2019dual} addressed this issue by separately modeling the audio and video sequences, while current methods focus more on considering cross-modal interactions when encoding visual and audio sequences. Works \cite{wu2019dual,xu2020cross,duan2021audio,xuan2020cross} mainly concentrate on segment-level feature encoding and alignment between different modalities to enhance the accuracy of event prediction. \cite{wu2019dual} is used for operational event-related functions. Zhou et al. \cite{zhou2021positive} proposed a positive sample propagation scheme by pruning weaker multi-modal interactions. Xuan et al. \cite{xuan2021discriminative,xuan2020cross} introduced a discriminative multi-modal attention module for sequential learning with a target function based on eigenvalues. Duan et al. \cite{duan2021audio} introduced joint collaborative learning with recurrent attention to aggregate multi-modal features. Lin and Wang \cite{lin2020audiovisual} introduced a transformer-based method that operates on groups of video frames based on audio-visual attention. Xu et al. \cite{xu2020cross} introduced multi-modal relation-aware audio-visual representation learning with an interaction module. 

\begin{figure*}[!h]
    \centering
    \includegraphics[width=\linewidth]{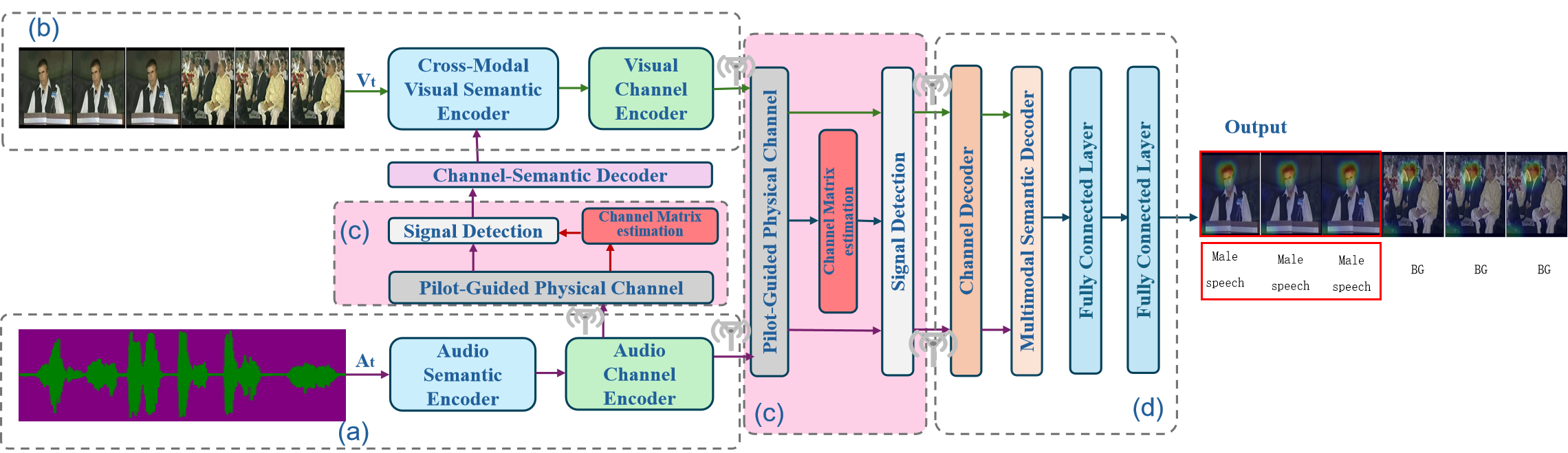}
 \caption{The overall framework of the pilot-guided mulimodal semantic communication for the multimodal task, which contains five modules: (a) the audio transmitter module; (b) the visual receiver-transmitter module; (c) the complex module of pilot-guided physical channel and signal detection; (d) the multimodal receiver module.}
    \label{fig:framework}
\end{figure*}


\section{Methodology}
In this paper, we propose a multimodal semantic communication system framework designed for multi-user terminals handling complex multimodal tasks. The framework includes a video transmitter with a single antenna and both an audio transmitter and a multimodal receiver with dual antennas. It addresses key challenges such as integrating and enhancing multimodal information, ensuring complementarity across users, and accurately decoding semantic information at the receiver end.

A major challenge in practice is the unknown nature of the channel state (channel matrix), which is crucial for signal detection and recovery. Without accurate knowledge of the channel matrix, the system may fail to recover transmitted information correctly. To tackle this issue, we introduce a pilot-guided channel matrix estimation module that accurately estimates the current channel state, enabling reliable signal detection and preventing communication failure.

\begin{definition}
\textbf{(Audio-Visual Event Localization, AVE)} involves identifying and categorizing segments within a video sequence that contain both audible and visible events. For a video sequence $S$ divided into $T$ one-second segments, with visual ($V=\{v_1,v_2,...,v_T\}$) and auditory ($A=\{a_1,a_2,...,a_T\}$) modalities, AVE aims to localize these audio-visual events and assign categories to each segment.

Formally, given a set of event categories $C$, including background, the task is to predict the event category for each segment $t$ as $y_t=\{y_{c,t}|y_{c,t}\in \{0,1\},c=1,2,...,C\}$, where $\sum_{c=1}^C y_{c,t}=1$. This involves classifying each segment as containing an audio-visual event or not, and if so, assigning it to one of the predefined categories.
\end{definition}

\subsection{Overall Framework}

The proposed pilot-guided multimodal semantic communication framework, illustrated in Fig.~\ref{fig:framework}, is designed to efficiently handle multimodal signals through five integral modules. The Audio Transmitter Module preprocesses and encodes audio signals using pilot signals to ensure high fidelity throughout the physical channel. The Visual Receiver-Transmitter Module manages video signals by compressing, encoding, transmitting, decoding, and reconstructing them to achieve high-quality video recovery. The Pilot-Guided Physical Channel and Signal Detection Module combines a pilot-guided channel model with signal detection algorithms to estimate the channel matrix, thereby improving signal recovery and decoding accuracy. The Multimodal Receiver Module processes and jointly decodes signals from both audio and visual sources, enhancing semantic understanding through the integration of multimodal data. Collectively, these modules create an integrated system that utilizes pilot-guided techniques to markedly improve the efficiency and reliability of multimodal semantic communication.

\subsection{Semantic Transmitter}

In this paper, the transmitter for multimodal semantic communication comprises several key components: First, it includes two modality-specific semantic encoders that extract semantic information from visual and audio inputs. Next, there is a pilot-based channel state estimation module, which estimates the channel matrix by guiding signal detection through pilot symbols. This estimation allows for accurate recovery of the transmitted information and helps prevent communication failures. Finally, channel encoders are used to protect information during transmission over the air.

\subsubsection{Semantic Encoders}
Given that visual information is captured in instantaneous frames, and auditory data sampling is segmented into discrete time intervals, exhibiting a discrete nature, we leverage this temporal characteristic. We segment the visual and auditory temporal data into continuous time segments to facilitate subsequent segment-based semantic feature extraction.

First, in each unimodal user transmitter, segment the video and audio time series \( S \) (corresponding to the same application scenario but sent to a remote receiver through different sensors and transmission ends) into \( T \) non-overlapping but consecutive segments, denoted \(\{S_t^v\}_{t=1}^T\) and \(\{S_t^a\}_{t=1}^T\). Each segment has a duration of one second. Here, \( S_t^v \) and \( S_t^a \) represent the visual and audio segments of the sequence, respectively.

\begin{figure}[!h]
    \centering
    \includegraphics[width=\linewidth]{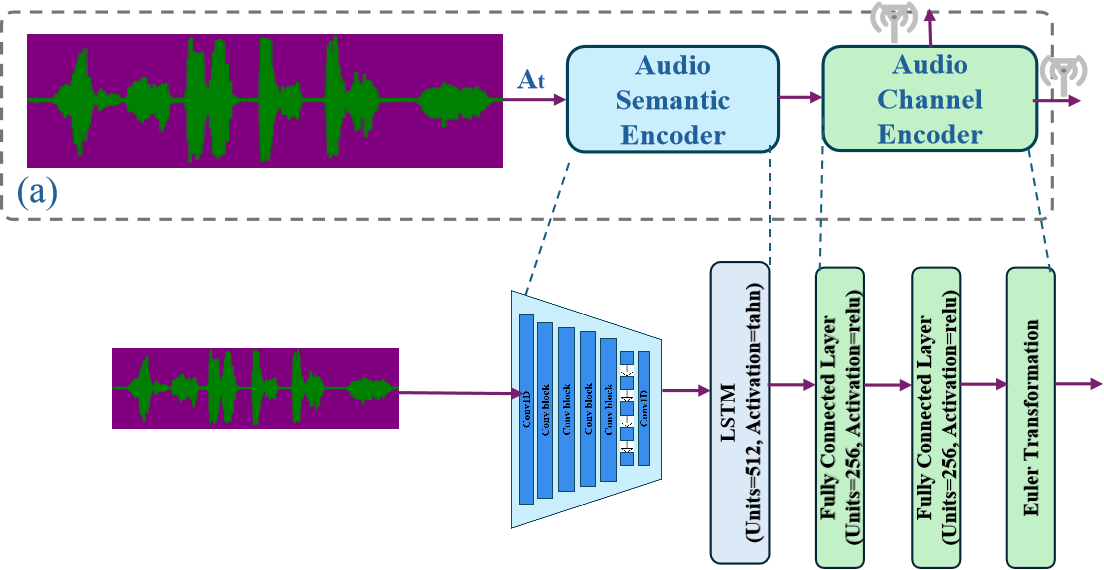}
 \caption{The audio transmitter module in Fig.~\ref{fig:framework}-(a).}
    \label{fig:frame-a}
\end{figure}

\noindent \textbf{For the audio semantic encoder:} As shown in Fig.~\ref{fig:frame-a}, given the audio segment sequence \(\{S_t^a\}_{t=1}^T\), we use the pre-trained VGG-19 model~\cite{VGG19} to extract audio features \(\mathcal{A} = \{\mathbf{a}_t\}_{t=1}^T \in \mathbb{R}^{T \times d_a}\). Here, \(d_a\) represents the feature dimension of the audio segments. Subsequently, an LSTM network with 512 units and \textit{tanh} activation is employed to perform semantic encoding of the audio source data. This semantic representation serves as the input for the channel encoding stage. 

\noindent \textbf{For the audio channel encoder:} The audio channel encoding aims to convert the semantic features into a format suitable for transmission over a communication channel. In this work, the channel encoding consists of two fully connected layers. The first fully connected layer reduces the dimensionality of the features, while the second fully connected layer reshapes the data for the channel.

\noindent \textbf{Euler Transformation:}  To transmit the semantic features via the channel, we propose the Euler transformation to transform the input from the original real vector space into a complex vector space. To maintain the consistency of the dot-product results before and after the transformation, we perform a split operation on a given token embedding $\mathbf{a}_t\in \mathbb{R}^{T \times d_a}$, transforming it into the complex vector space. Based on Euler's Formula\footnote{Euler's Formula can be written as: $\mathbf{r}_{\mathbf{a}_t}+i\mathbf{s}_{\mathbf{a}_t}\text{(rectangular form)}=\lambda e^{i\theta_{\mathbf{a}_t}}\text{(polar form)}$\cite{tian2023eulernet}.}, we obtain its polar-form representation as the complex signal:
\begin{equation}
    \tilde{\mathbf{a}_t}=\mathcal{F}(\mathbf{a}_t)=\mathcal{F}([\mathbf{r}_{\mathbf{a}_t},\mathbf{s}_{\mathbf{a}_t}])=\lambda_{\mathbf{a}_t}\exp(i\theta_{\mathbf{a}_t}),
\end{equation}
where $\mathbf{r}_{\mathbf{a}_t}=\mathbf{a}_t[1:\frac{d_a}{2}]$ and $\mathbf{s}_{\mathbf{a}_t}=\mathbf{a}_t[\frac{d_a}{2}+1:d_a]$ are the two vectors obtained after splitting $\mathbf{a}_t$, serving as the real and imaginary parts of the transformed complex vector $\tilde{\mathbf{a}_t}$, respectively. Here, $\lambda_{\mathbf{a}_t}=\sqrt{\mathbf{r}_{\mathbf{a}_t}^2+\mathbf{s}_{\mathbf{a}_t}^2}$ and $\theta_{\mathbf{a}_t}=\text{atan2}(\mathbf{s}_{\mathbf{a}_t},\mathbf{r}_{\mathbf{a}_t})$ are the polar-form representations of $\tilde{\mathbf{a}_t}$ using Euler’s formula.

\begin{figure}[!h]
    \centering
    \includegraphics[width=\linewidth]{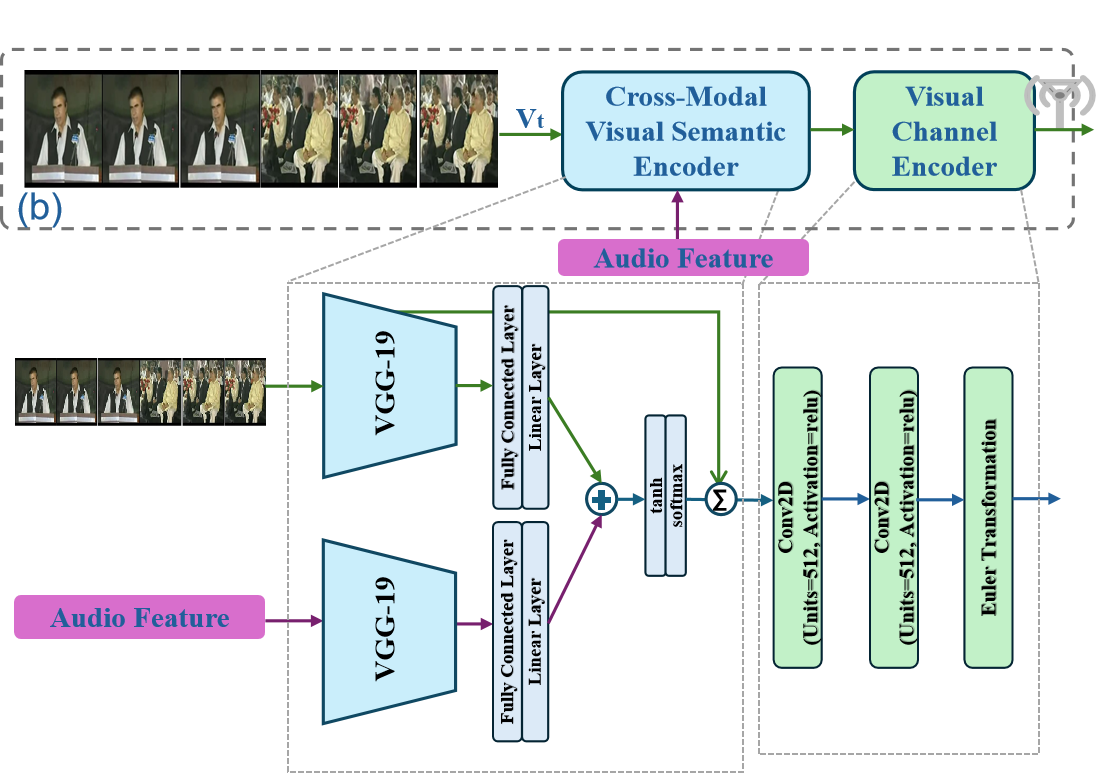}
 \caption{The visual receiver-transmitter module in Fig.~\ref{fig:framework}-(b).}
    \label{fig:frame-b}
\end{figure}
\noindent \textbf{For the visual semantic encoder}: As shown in Fig.~\ref{fig:frame-b}, given the video segment sequence \(\{S_t^v\}_{t=1}^T\), we use the pre-trained VGG-19 model to extract visual features \(\mathcal{V} = \{\mathbf{v}_t\}_{t=1}^T \in \mathbb{R}^{T \times k \times d_v}\). Here, \(d_v\) represents the feature dimension of the visual segments, and \(k\) represents the vectorized spatial dimension of each feature map. Initially, we utilize AGVA~\cite{ave} to encode visual features \(\mathcal{V}^{att} = \{\mathbf{v}_t^{att}\}_{t=1}^T \in \mathbb{R}^{T \times d_v}\), guiding the model's focus towards visual regions closely correlated with the audio content. At each time step \(t\), \(\mathbf{v}_t^{att}\) is computed by:
\begin{equation}
    \mathbf{v}_t^{att} = f_{att}(\mathbf{v}_t, \mathbf{a}^{*}_t) = \alpha_t \mathbf{v}_t,
\end{equation}
where \(\mathbf{a}^{*}_t\) is the audio feature information received via the channel from the user end with the audio source, and \(\alpha_t\) denotes the attention weight vector, which can be computed using an MLP with a Softmax activation function:
\begin{equation}
    \alpha_t = \sigma\left(W_f \tanh(W_v^1 M_v (\mathbf{v}_t) + W_a^1 M_a (\mathbf{a}^{*}_t))\right),
\end{equation}
where \(M_v\) and \(M_a\) are transformation functions achieved via a non-linear dense layer, projecting audio and visual features into a shared dimension \(d\), and \(W_v^1 \in \mathbb{R}^{k \times d}\), \(W_a^1 \in \mathbb{R}^{k \times d}\), and \(W_f \in \mathbb{R}^{1 \times k}\) are trainable parameters. The function \(\sigma(\cdot)\) represents a Softmax activation function.

\noindent \textbf{For the visual channel encoder}: Similar to the structure of the audio channel encoder, the visual channel encoder consists of two Conv2 blocks and an Euler Transformation block. Its purpose is to convert the semantic features into a format suitable for transmission over a communication channel. The representation of the visual channel encoder is given by:
\begin{equation}
    \widetilde{\mathbf{v}_t^{att}} = \mathcal{F}(\mathbf{v}_t^{att}) = \mathcal{F}([\mathbf{r}_{\mathbf{v}_t^{att}}, \mathbf{s}_{\mathbf{v}_t^{att}}]) = \lambda_{\mathbf{v}_t^{att}} \exp(i\theta_{\mathbf{v}_t^{att}}),
\end{equation}
where \(\mathbf{r}_{\mathbf{v}_t^{att}} = \mathbf{v}_t^{att}[1:\frac{d_v}{2}]\) and \(\mathbf{s}_{\mathbf{v}_t^{att}} = \mathbf{v}_t^{att}[\frac{d_v}{2}+1:d_v]\) are the two vectors obtained after splitting \(\mathbf{v}_t^{att}\), serving as the real and imaginary parts of the transformed complex vector \(\widetilde{\mathbf{v}_t^{att}}\), respectively. Here, \(\lambda_{\mathbf{v}_t^{att}} = \sqrt{\mathbf{r}_{\mathbf{v}_t^{att}}^2 + \mathbf{s}_{\mathbf{v}_t^{att}}^2}\) and \(\theta_{\mathbf{v}_t^{att}} = \text{atan2}(\mathbf{s}_{\mathbf{v}_t^{att}}, \mathbf{r}_{\mathbf{v}_t^{att}})\) are the polar-form representations of \(\widetilde{\mathbf{v}_t^{att}}\).

\subsubsection{Pilot-guided Physical Channel and Signal Detection} 
In wireless communication systems, dynamic changes in channel characteristics—varying with time, frequency, and space—cause signal attenuation and phase shifts. To address these variations, pilot signals are added before signal transmission, allowing the receiver to accurately estimate current channel conditions, simplify system design, and enhance robustness. Pilot signals enable the receiver to estimate channel state (including attenuation and phase shift), simplify signal processing without complex algorithms, distinguish between multiple users to reduce interference, and provide sufficient information for channel estimation even under poor conditions.
\vspace{-0.5cm}
\begin{figure}[!h]
    \centering
    \includegraphics[width=\linewidth]{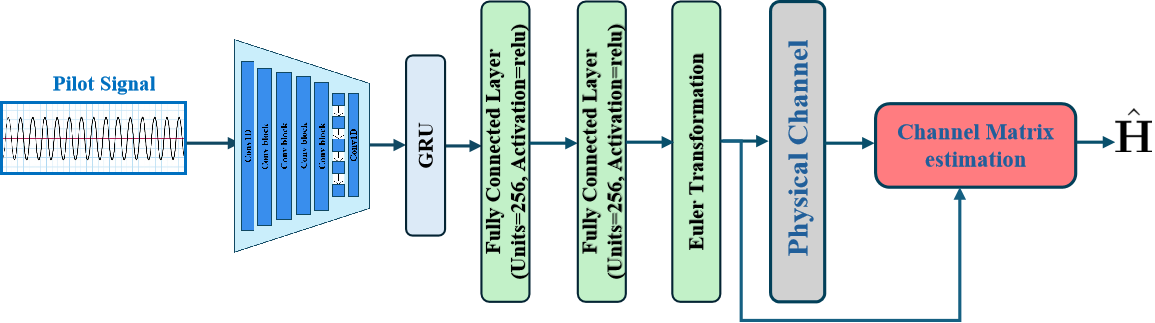}\vspace{-0.45cm}
 \caption{The complex module of pilot-guided physical channel and signal detection in Fig.~\ref{fig:framework}-(c).}
    \label{fig:fame-b}
\end{figure}

\noindent \textbf{Pilot-guided Channel Estimation:}

The pilot-guided channel estimation process, as illustrated in Fig.~\ref{fig:fame-b}, is divided into two main stages: source-channel encoding and channel matrix estimation.

In the source-channel encoding stage, we start with feature extraction and encoding. This involves applying a 1D convolution with \( C \) channels and a kernel size of 7, followed by \( B \) convolution blocks. Each block consists of a residual unit, which includes two convolutions with kernel size 3 and a skip connection, and is followed by a downsampling layer with a strided convolution, where the kernel size \( K \) is twice the stride \( S \). The number of channels doubles at each downsampling step. After these layers, a final 1D convolution layer with a kernel size of 7 and \( D \) output channels is applied. Subsequently, a Gated Recurrent Unit (GRU) layer processes the output from the convolution layers, and the result is further abstracted through fully connected layers with ReLU activation functions. Finally, the output is transformed into a complex representation via the Euler transformation to prepare it for channel estimation.

The channel matrix estimation stage aims to estimate the channel response based on the known transmitted signal and the received signal. For this purpose, we use the Least Squares (LS) method. The pilot signal used is a 10-second sinusoidal wave encoded by the audio source-channel encoder. After transmission, both the received signal \( y \) and the transmitted signal \( x \) are complex and span multiple time steps.

For each time step \( t \), the channel response estimate \( \hat{\mathbf{H}}_t \) is calculated using the LS method:
\begin{equation}
    \hat{\mathbf{H}}_t = (\mathbf{x}_t^H \mathbf{x}_t)^{-1} \mathbf{x}_t^H y_t.
\end{equation}

To determine the best estimate, we compute the squared error for each time step and identify \( t_{\text{min}} \), the time step with the smallest squared error:
\begin{equation}
   t_{\text{min}} = \arg \min_t \| \hat{\mathbf{H}}_t \mathbf{x}_t - y_t \|^2. 
\end{equation}

The channel response estimate with the minimum squared error is then selected as: $\hat{H}_{\text{best}} = \hat{H}_{t_{\text{min}}}$. This method ensures that the channel coefficient matrix is accurately determined by selecting the estimate that minimizes the squared error across all time steps.

\noindent \textbf{Signal Detection:}

Unlike existing semantic communication studies that typically assume a known and unchanging channel state (referred to as a perfect channel with a known channel matrix), this paper addresses the practical scenario where the channel matrix is unknown (blind). Existing research often uses this perfect channel matrix directly for signal detection. In contrast, we utilize the estimated channel matrix \(\hat{\mathbf{H}}\) obtained from \textit{channel matrix estimation} for signal detection. By leveraging additional domain knowledge, i.e., channel estimation, we can improve the training speed and enhance the final decision accuracy. With the channel gain and zero-forcing detector, the transmitted signal can be estimated by:

\begin{equation}
    \hat{\mathbf{X}} = (\hat{\mathbf{H}}^H \hat{\mathbf{H}})^{-1} \hat{\mathbf{H}}^H \mathbf{Y} = \mathbf{X} + \hat{\mathbf{N}},
    \label{eq:signeldetect}
\end{equation}
where \(\hat{\mathbf{X}} = [\hat{\mathcal{A}}^*, \hat{\mathcal{V}}]\) is the estimated information for the audio and video users, and \(\hat{\mathbf{N}} = (\hat{\mathbf{H}}^H \hat{\mathbf{H}})^{-1} \hat{\mathbf{H}}^H \mathbf{N}\) represents the impact of noise. This operation in Eq.~\ref{eq:signeldetect} transforms the channel effect from multiplicative noise to additive noise, significantly reducing the learning burden.

\subsection{Semantic Receiver}
In the multimodal semantic communication system proposed in this paper, there are two types of receivers: the visual user terminal receiver (Single-Input Single-Output, SISO) and the multimodal receiver (Multiple-Input Multiple-Output, MIMO). 

For the physical channel from the audio terminal to the video terminal (with SISO), the received signal is represented as \( \mathbf{Y} \in \mathbb{C}^{L_c} \), where \( L_c \) denotes the length of the channel. For the physical channel in the multimodal receiver (with MIMO), the received signal is represented as \( \mathbf{Y} \in \mathbb{C}^{M \times L_c} \), where \( M \) is the number of antennas at the receiver and \( L_c \) is the length of the channel. In both cases, the received signal can be expressed by the equation: $\mathbf{Y} = \mathbf{H} \mathbf{X} + \mathbf{N}$, where \( \mathbf{H} \) denotes the channel matrix, \( \mathbf{X} \) is the transmitted signal, and \( \mathbf{N} \) represents the noise matrix.

In the case of the multimodal receiver physical channel, \( \mathbf{X}^T = [x_1, x_2, \ldots, x_K] \in \mathbb{C}^{L_c \times K} \) represents the transmit symbols from all \( K \) users, and \( \mathbf{H} = [\mathbf{h}_1, \mathbf{h}_2, \ldots, \mathbf{h}_K] \in \mathbb{C}^{L} \) is the channel matrix between the base station (BS) and the users. This study develops models for three types of noisy wireless channels: additive white Gaussian noise (AWGN) channel, slow Rayleigh fading channel with AWGN, and Rician fading channel.

\subsubsection{Channel Decoders}

After signal detection, the estimated complex signals are first reshaped to the size suitable for the subsequent neural networks using a reshape layer. Then, the signals are semantically recovered by the channel decoders for audio and video. Both channel decoders must first convert the complex signals to real values. To handle the Euler Transformation applied during channel encoding at the transmitter, we perform \textit{Complex Decomposition}. Specifically, we conduct the inverse transformation (i.e., \(\mathcal{F}^{-1}\)) to convert the complex signals into the real vector space, as shown by:
\begin{align}
     \mathbf{a}_t &= \mathcal{F}^{-1}(\tilde{\mathbf{a}}_t)\nonumber\\ 
     &= \mathcal{F}^{-1}(\lambda'_a \exp(i\theta_{\mathbf{a}_t}')) \nonumber\\&= [\lambda'_a \cos(\theta_{\mathbf{a}_t}'); \lambda_a' \sin(\theta_{\mathbf{a}_t}')];\\
     \mathbf{v}_t^{att} &= \mathcal{F}^{-1}( \widetilde{\mathbf{v}_t^{att}})\nonumber\\
     &= \mathcal{F}^{-1}(\lambda'_v \exp(i\theta_{\mathbf{v}_t^{att}}')) \nonumber\\&= [\lambda'_v \cos(\theta_{\mathbf{v}_t^{att}}'); \lambda'_v \sin(\theta_{\mathbf{v}_t^{att}}')],
\end{align}
where \([;]\) denotes the concatenation operation.
\begin{figure}[!h]
    \centering
    \includegraphics[width=\linewidth]{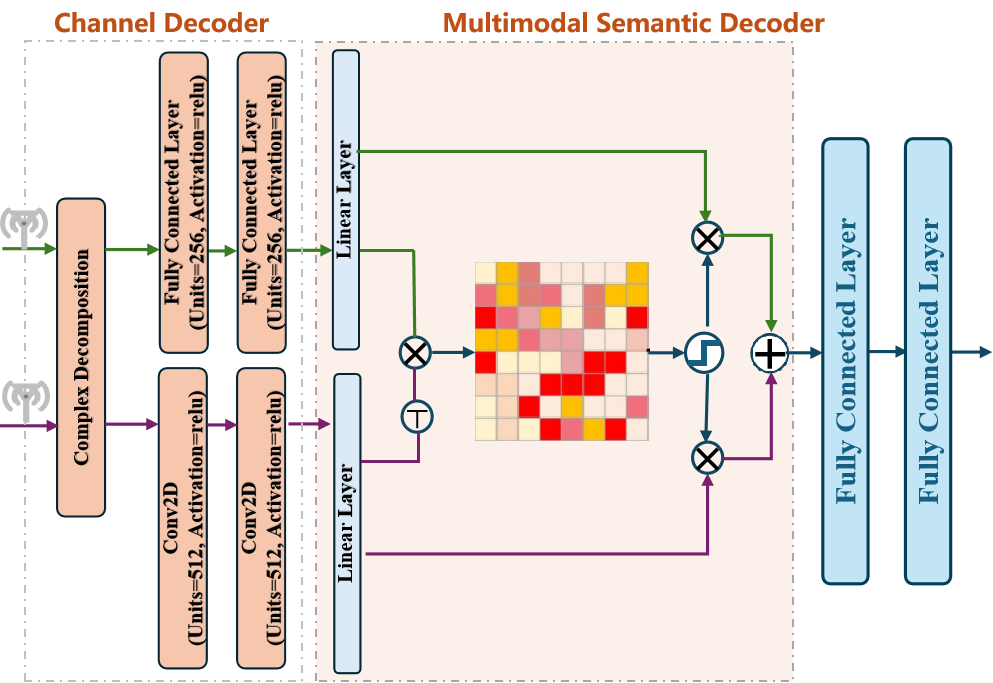}
    \vspace{-0.5cm}
 \caption{The multimodal receiver module in Fig.~\ref{fig:framework}-(d).}
    \label{fig:frame-d}
\end{figure}

As shown in Fig.~\ref{fig:frame-d}, the complex decomposition block recovers the real-valued representations of the received multimodal features. To handle the semantic decoding block for multimodal tasks, the channel decoding section processes visual and audio features as follows: For the audio channel decoder, the received audio is processed through two fully connected layers (units=256, activation='ReLU'). For the visual channel decoder, the received video features are processed through two Conv2D blocks (units=512, activation='ReLU'). To ensure compatibility for the multimodal semantic decoding block, both audio and visual features are mapped to the same feature dimension space using a linear layer: $\textbf{M}^a,\textbf{M}^v$.

\subsubsection{Semantic Decoder}
For the multimodal task of Audio-Visual Event Localization, we utilize the \textit{Positive Sample Propagation}~\cite{psp} as the semantic decoder to merge audio and visual semantic information. A correlation-based scoring matrix identifies temporal intervals with strong connections among audio-visual segments. The connection strength between audio-visual segments \((\mathbf{M}^v, \mathbf{M}^a)\) is calculated as follows:
\begin{equation}
\omega^{va} = \frac{\left(\mathbf{M}^v W_v^2\right)\left(\mathbf{M}^a W_a^2\right)^\top}{\sqrt{d_l}},
\label{eq:correlation}
\end{equation}
where \(W_v^2\in \mathbb{R}^{d_l \times d_s}\) and \(W_a^2 \in \mathbb{R}^{d_l \times d_s}\) are trained parameters, with \(d_l\) representing the input feature dimension. \(\omega^{va}\) and \(\omega^{av} \in \mathbb{R}^{T \times T}\) are similarity matrices for audio-visual segments.

To retain segments with strong connections, we filter out negative and weak connections using the Rectified Linear Unit (ReLU) activation function and row-wise \(\ell_1\) normalization to obtain the normalized similarity matrix \(\hat{\omega}^{va}\):
\begin{equation}
\hat{\omega}^{va} = \textit{LayerNorm}(\textit{ReLU}(\omega^{va})),
\end{equation}
where \(\textit{LayerNorm}(\cdot)\) denotes row-wise \(\ell_1\) normalization, and \(\textit{ReLU}(\cdot)\) is the ReLU activation function.

Weak connections are filtered out using a threshold as follows:
\begin{align}
\eta^{va} &= \textit{LayerNorm}(\hat{\omega}^{va} \mathbb{I}\{\hat{\omega}^{va} > \tau_{1}\}), \label{eq:similar1} \\
\eta^{av} &= \textit{LayerNorm}(\hat{\omega}^{av} \mathbb{I}\{\hat{\omega}^{av} > \tau_{1}\}),
\label{eq:similar2}
\end{align}
where \(\tau_{1}\) is a threshold hyperparameter, and \(\mathbb{I}\{\cdot\}\) is a binary function that outputs 1 if the condition is met and 0 otherwise. Row-wise \(\ell_1\) normalization is applied again to generate the final similarity matrices \(\eta^{va}\) and \(\eta^{av} \in \mathbb{R}^{T \times T}\), ensuring that the matrices are appropriately scaled and retain essential relational information.

Based on the positive sample propagation \(\eta^{va}\) (\(\eta^{av}\)), we obtain the multimodal semantic features for the AVE task after semantic decoding as follows:
\begin{align}
    &\mathbf{SD}_a = \eta^{av}(\mathbf{M}^v \mathbf{W}^a_2) + \mathbf{M}^a, \quad \mathbf{SD}_a \in \mathbb{R}^{T \times d_l},\\
    &\mathbf{SD}_v = \eta^{va}(\mathbf{M}^a \mathbf{W}^v_2) + \mathbf{M}^v, \quad \mathbf{SD}_v \in \mathbb{R}^{T \times d_l},
\end{align}
where \(\mathbf{W}^a_2\) and \(\mathbf{W}^v_2\) are parameters defining linear transformations, and both belong to \(\mathbb{R}^{d_l \times d_l}\).

\subsection{Training Strategy}
In this paper, the multimodal task (Audio-Visual Event Localization) is framed as a classification problem. We use a neural network trained with a widely used classification loss function, the cross-entropy loss.

In the multimodal receiver module, as shown in Figure~\ref{fig:frame-d}, the multimodal semantic decoding process includes semantic correlation segmentation extraction. Here, we align intervals to obtain highly similar audio (visual) content to the given visual (audio) content, thereby updating audio and visual features. These features are transformed into the same embedding space using linear layers and then merged into fused features through simple averaging. The fused features are further processed using two fully connected layers, and the softmax function is applied to obtain the predicted event labels $X \in \mathbb{R}^{T \times C}$.

The cross-entropy (CE) loss is employed as the objective function based on the predicted event labels $X$ and the ground truth $Y$, expressed as:
\begin{equation}
    \mathcal{L}_{cls} = -\frac{1}{TC} \sum_{t=1}^{T} \sum_{c=1}^{C} Y_{tc}\log{(X_{tc})},
    \label{cls}
\end{equation}
allowing the model to predict the event category of video segments.

In addition to classification loss, we introduce an audio-visual segment similarity loss, encouraging the model to learn highly correlated features of audio and visual segments when they share the same event label. For a video with $T$ segments, a label vector $G = \{g_{t} \in \{0,1\} \mid t=1,2, \dots, T\} \in \mathbb{R}^{1 \times T}$ is defined, and cosine similarity $S_{cos}$ is calculated between visual and audio features after $\ell_1$ normalization. The proposed loss is then given by:
\begin{equation}
    \mathcal{L}_{avs} = \mathcal{L}_{MSE}(S_{cos}, G),
    \label{avs}
\end{equation}
where $\mathcal{L}_{MSE}(\cdot, \cdot)$ represents the mean squared error (MSE) between two vectors.

Combining Eq.~\ref{cls} and Eq.~\ref{avs}, the overall objective function for the fully-supervised setting $\mathcal{L}_{total}$ can be computed as:
\begin{equation}
    \mathcal{L}_{total} = \mathcal{L}_{cls} + \beta \mathcal{L}_{avs},
\end{equation}
where $\beta$ is a hyper-parameter to balance the two losses.

Additionally, we utilize the Signal-to-Noise Ratio (SNR) to assess the quality of the communication channel $SNR=10\log10\frac{P}{\sigma^2}$,
where \( P \) denotes the average power of the channel's input signal, while \( \sigma^2 \) represents the noise power. The unit of SNR is decibels (dB), which measures the signal strength relative to the background noise.

To enhance the system's noise resistance, we conduct joint training on the transmitters and receiver. Specifically, we consider that the far-end server (base station) has a higher transmission power than the user equipment. As a result, during both training and testing phases, we set a range of SNR values for the feedback branch $\{0dB, 3dB,6dB,9dB, 12dB,15dB,18dB, 21dB, 24dB,27dB,$\\$30dB\}$. These SNR values help to simulate various noise conditions, thereby evaluating and optimizing the system's performance under different environments.

\section{Experiments}
\begin{figure*}[!t]
    \centering
    \includegraphics[width=\linewidth]{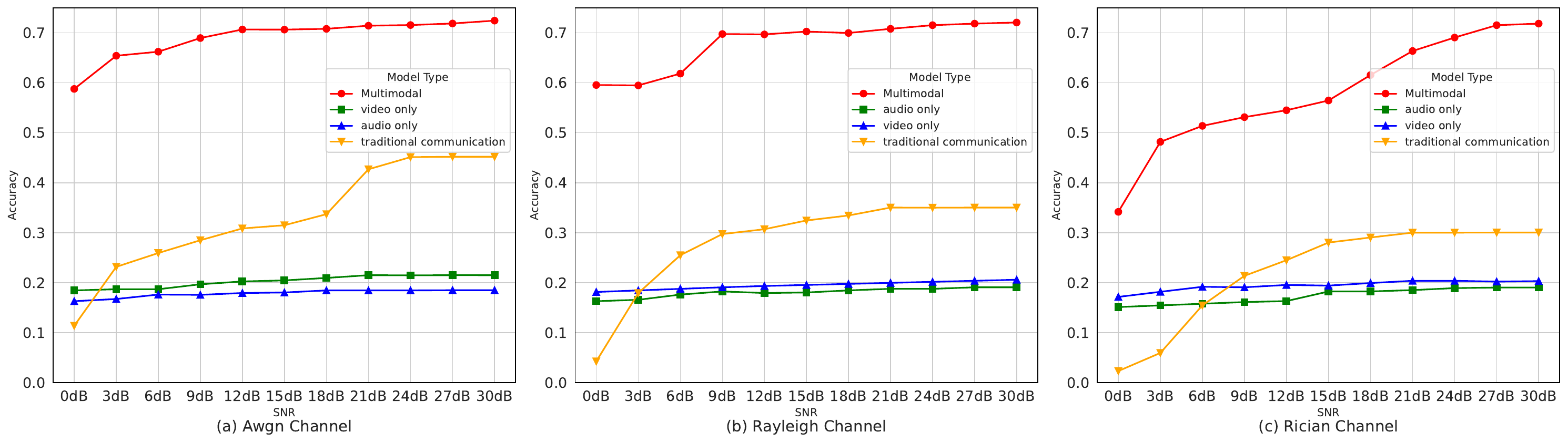}
    \vspace{-0.8cm}
    \caption{The accuracy for various testing channels based on different trained models (for RQ1).}
    \label{fig:ex_snr}
\end{figure*}

The experiments are designed to investigate the research questions:
\begin{itemize}
    \item RQ1. How does our semantic communication method compare to traditional source coding and channel coding methods over different channels with unknown CSI? (Section~\ref{sec:ex_withCSI})
    \item RQ2. How to validate the effectiveness of the proposed multimodal semantic encoding in the multimodal task (AVE)? (Section~\ref{sec:ex_SED})
    \item RQ3. How to validate the effectiveness of the pilot-guided channel state estimation module? (Section~\ref{sec:ex_woCSI})
\end{itemize}

\subsection{Dataset Description}
In this work, we use the Audio-Visual Event (AVE) subset of the Audioset dataset~\cite{tian2018audio}, widely employed in Audio-Visual Event (AVE) tasks. The Audioset contains 4,143 videos categorized into 28 distinct event types with temporal annotations for audio-visual events. Each video contains at least one audio-visual event lasting a minimum of 2 seconds, spanning diverse domains such as human and animal activities, musical performances, and vehicle sounds. Videos are distributed across categories, with each containing between 60 to 188 videos. More than half of the videos feature audiovisual events that span the entire 10-second duration. Our primary task involves event localization, aiming to identify the event type for each one-second segment in synchronized video and audio streams, thereby accurately localizing events within the video.

\subsection{Evaluation Metrics}

In this paper, we primarily focus on the multimodal task of audio-visual event localization, which is a classification task. We will use accuracy as the main evaluation metric, a widely adopted measure in existing research. The category label of each segment is predicted in a fully supervised setting. Following the existing work~\cite{lin2019dual,xu2020cross,xuan2020cross,ave,psp}, we adopt the classification accuracy of each segment as the evaluation metric. Additionally, this paper evaluates performance based on different Signal-to-Noise Ratios (SNR) and various channel states, using accuracy as the primary evaluation metric.

\subsection{Implementation Details}
In line with the original AVE study work~\cite{ave,psp}, we utilize the same visual and audio features to ensure a fair comparison. For visual feature extraction, we use VGG-19~\cite{VGG19} pre-trained on ImageNet~\cite{krizhevsky2012imagenet}. For audio features, we first convert raw audio into log-mel spectrograms and then use a VGG-like network~\cite{hershey2017cnn} pre-trained on AudioSet~\cite{gemmeke2017audio}.  

The framework is trained using the ScheduledOptim optimizer with an initial learning rate of $3e\text{-}4$, which decreases by a factor of $0.1$ every $10$ epochs. The model is trained for $300$ epochs with a batch size of $64$. Additionally, based on extensive experiments and analysis, we set $\lambda = 100$ and $\tau_1 = 0.099$. All experiments are conducted on a single NVIDIA RTX 2080Ti GPU, covering both training and evaluation processes. Our method's code is developed based on \url{https://github.com/dimlight13/MU_SC_for_VQA}.

\subsection{Experimental Results and Analysis}
\subsubsection{The performance of our methods with channel matrix estimation}
\label{sec:ex_withCSI}

\begin{figure*}[!h]
    \centering
    \includegraphics[width=0.9\linewidth]{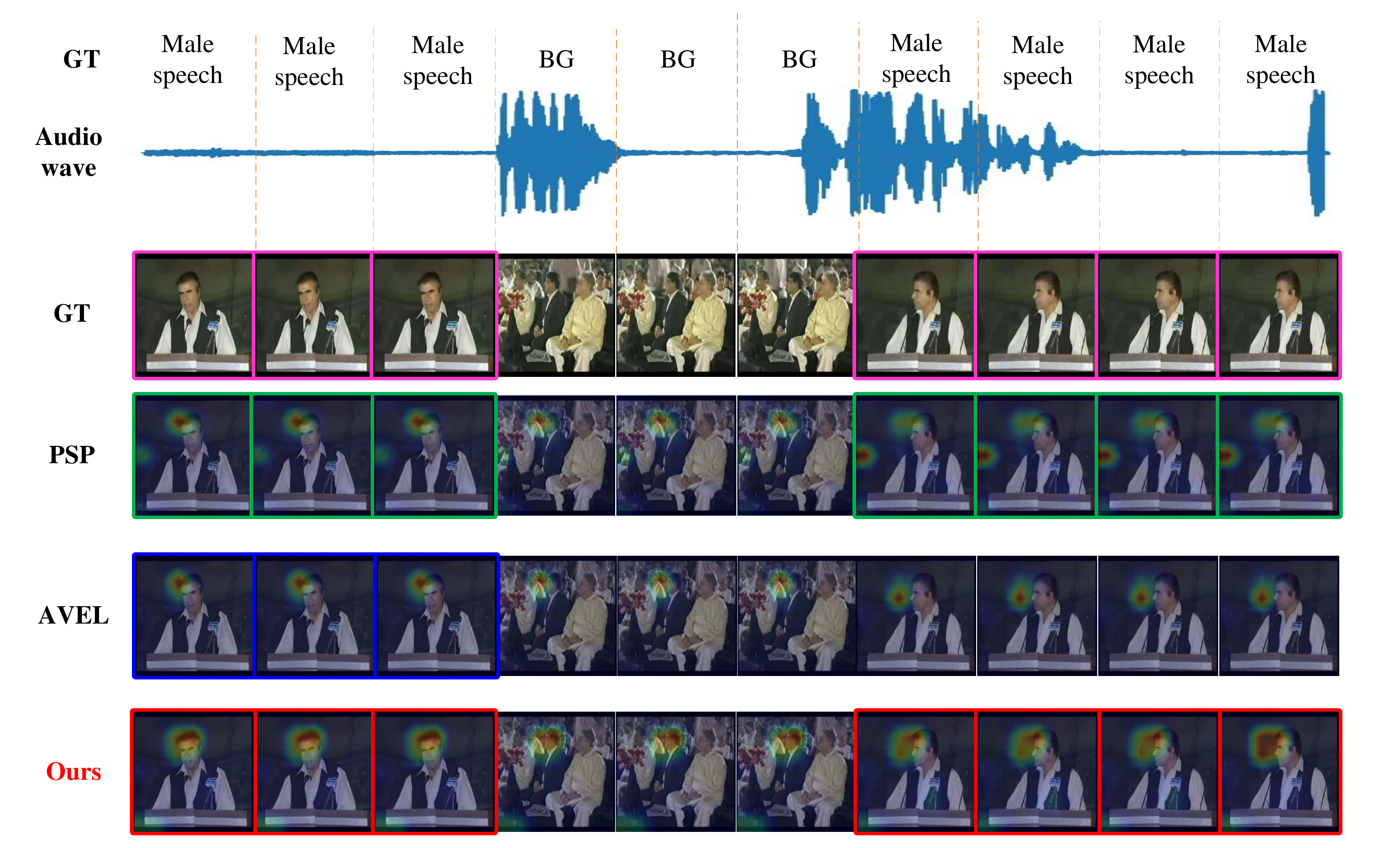}
    \vspace{-0.45cm}
 \caption{The qualitative example of AVE localization. For the video, only the first three segments contain the visual and audio signals of the event male speech. The pink boxes represent ground truth labels. The blue and green boxes indicate predictions of baselines (error-free transmission), and the red boxes indicate the predictions of ours, respectively. Besides, we visualize the attention effect on the images. It is clear that our method produces more accurate localization and that our attended regions better overlap with the sound sources (for RQ2).}
    \label{fig:hotmap}
\end{figure*}


We evaluated our proposed multimodal semantic communication framework against two baseline scenarios: (1) classification using only video or audio, and (2) traditional communication methods. The traditional approach involved separate source and channel coding: converting video into frames with JPEG compression and using pulse-code modulation (PCM) for audio, with LDPC codes for channel coding.

\noindent \textbf{Performance Comparison}: Fig.~\ref{fig:ex_snr} shows the relationship between localization accuracy and Signal-to-Noise Ratio (SNR) across various channels—Additive White Gaussian Noise (AWGN), Rayleigh fading, and Rician fading. The results highlight the following:
\begin{itemize}
    \item Single-Modality Methods: These methods, which rely on either video or audio alone, exhibit lower accuracy due to missing modality information and lack of cross-modal guidance.
 \item Traditional Methods: Perform similarly to single-modality methods but with limited effectiveness in handling complex channel conditions.
 \item Proposed Multimodal Approach: Significantly outperforms single-modality and traditional methods. It achieves near-optimal localization accuracy under high SNR conditions and maintains robust performance even under low SNR and multipath fading scenarios. 
\end{itemize}

\begin{table*}[!t]
\centering
\caption{Performance Comparison of Single-Modal and Multimodal Semantic Transmission Under Different SNRs with and Without Channel Matrix Estimation (for RQ3).}
\resizebox{\textwidth}{!}{
\begin{tabular}{ccccccccccccc} 
\hline
\multirow{2}{*}{\textbf{Model Type}} & \multicolumn{11}{c}{\textbf{SNR}} \\\cmidrule(lr){2-12} 
& \textbf{0dB} & \textbf{3dB} & \textbf{6dB} & \textbf{9dB} & \textbf{12dB} & \textbf{15dB} & \textbf{18dB} & \textbf{21dB} & \textbf{24dB} & \textbf{27dB} & \textbf{30dB} \\\hline
\multicolumn{12}{c}{\underline{\textbf{AWGN Channel}}}\\ %
\textbf{Audio (only)} & 0.1535 & 0.1612 & 0.1602 & 0.1648 & 0.1740 &0.1740 & 0.1727 & 0.1708&0.1776 & \underline{\textbf{0.1867}} & 0.1808 \\
&\textcolor{blue}{$\downarrow$} \textcolor{blue}{5.69\%} & \textcolor{blue}{$\downarrow$} \textcolor{blue}{3.58\%} & \textcolor{blue}{$\downarrow$} \textcolor{blue}{9.00\%} & \textcolor{blue}{$\downarrow$} \textcolor{blue}{6.08\%} & \textcolor{blue}{$\downarrow$} \textcolor{blue}{2.88\%} & \textcolor{blue}{$\downarrow$} \textcolor{blue}{3.47\%} & \textcolor{blue}{$\downarrow$} \textcolor{blue}{6.36\%} & \textcolor{blue}{$\downarrow$} \textcolor{blue}{7.34\%} & \textcolor{blue}{$\downarrow$} \textcolor{blue}{3.67\%} &  & \textcolor{blue}{$\downarrow$} \textcolor{blue}{2.13\%}\\
\textbf{Video (only)} &0.1844 & 0.1844 & 0.1864 & 0.1855 & 0.2016 & 0.2055 & 0.2016 & \underline{\textbf{0.2094}} & 0.2000 & 0.2068 & 0.2050 \\
&& \textcolor{blue}{$\downarrow$}\textcolor{blue}{1.26\%} & \textcolor{blue}{$\downarrow$} \textcolor{blue}{0.20\%} & \textcolor{blue}{$\downarrow$} \textcolor{blue}{5.66\%} & \textcolor{blue}{$\downarrow$} \textcolor{blue}{0.39\%} &  & \textcolor{blue}{$\downarrow$} \textcolor{blue}{3.73\%} & \textcolor{blue}{$\downarrow$} \textcolor{blue}{2.55\%} & \textcolor{blue}{$\downarrow$} \textcolor{blue}{6.77\%}& \textcolor{blue}{$\downarrow$} \textcolor{blue}{3.76\%} & \textcolor{blue}{$\downarrow$} \textcolor{blue}{4.57\%}\\
\textbf{Multimodal}& 0.2341 & 0.5380& 0.5865 & 0.5898 & 0.6005 & 0.6000& 0.5711 & 0.6143 & 0.6120 & \underline{\textbf{0.6177}} & 0.6144 \\
&\textcolor{blue}{$\downarrow$} \textcolor{blue}{60.17\%} & \textcolor{blue}{$\downarrow$} \textcolor{blue}{17.75\%} & \textcolor{blue}{$\downarrow$} \textcolor{blue}{11.44\%} & \textcolor{blue}{$\downarrow$} \textcolor{blue}{14.46\%} & \textcolor{blue}{$\downarrow$} \textcolor{blue}{15.03\%} & \textcolor{blue}{$\downarrow$} \textcolor{blue}{15.08\%} & \textcolor{blue}{$\downarrow$} \textcolor{blue}{19.35\%} & \textcolor{blue}{$\downarrow$} \textcolor{blue}{14.00\%} & \textcolor{blue}{$\downarrow$} \textcolor{blue}{14.48\%} & \textcolor{blue}{$\downarrow$} \textcolor{blue}{14.06\%} & \textcolor{blue}{$\downarrow$} \textcolor{blue}{15.20\%} 
\\\hline
\multicolumn{12}{c}{\underline{\textbf{Rayleigh Channel}}}\\
\textbf{Audio (only)}&0.1429 & 0.1703 & 0.1698 & 0.1778 & 0.1792 & 0.1841 & 0.1799 & 0.1854 & 0.1901 & \underline{\textbf{0.1922}} & 0.1914 \\%
&\textcolor{blue}{$\downarrow$} \textcolor{blue}{{12.19\%}} & & \textcolor{blue}{$\downarrow$} \textcolor{blue}{{3.52\%}} & \textcolor{blue}{$\downarrow$} \textcolor{blue}{{2.49\%}} & &  & \textcolor{blue}{$\downarrow$} \textcolor{blue}{{2.40\%}} & \textcolor{blue}{$\downarrow$} \textcolor{blue}{{1.17\%}} &  &  &  \\
\textbf{Video (only)}& 0.1705 & 0.1721 & 0.1797 & 0.1880 & 0.1836 & 0.1911 & 0.1986 & 0.1912 & \underline{\textbf{0.2016}} & 0.2000 & 0.1976\\
& \textcolor{blue}{$\downarrow$} \textcolor{blue}{5.93\%} & \textcolor{blue}{$\downarrow$} \textcolor{blue}{6.64\%} & \textcolor{blue}{$\downarrow$} \textcolor{blue}{4.17\%} & \textcolor{blue}{$\downarrow$} \textcolor{blue}{1.37\%} & \textcolor{blue}{$\downarrow$} \textcolor{blue}{4.99\%} & \textcolor{blue}{$\downarrow$} \textcolor{blue}{2.13\%} &  & \textcolor{blue}{$\downarrow$} \textcolor{blue}{4.13\%} & & \textcolor{blue}{$\downarrow$} \textcolor{blue}{1.79\%} & \textcolor{blue}{$\downarrow$} \textcolor{blue}{3.95\%} \\
\textbf{Multimodal} & 0.1844& 0.1836& 0.2154& 0.2284 & 0.2448 & 0.2826 & 0.2932& 0.3169& 0.2932 & \underline{\textbf{0.3487}} & 0.3432 \\
& \textcolor{blue}{$\downarrow$} \textcolor{blue}{69.04\%} & \textcolor{blue}{$\downarrow$} \textcolor{blue}{69.13\%} & \textcolor{blue}{$\downarrow$} \textcolor{blue}{65.18\%} & \textcolor{blue}{$\downarrow$} \textcolor{blue}{67.26\%} & \textcolor{blue}{$\downarrow$} \textcolor{blue}{64.87\%} & \textcolor{blue}{$\downarrow$} \textcolor{blue}{59.79\%} & \textcolor{blue}{$\downarrow$} \textcolor{blue}{58.10\%} & \textcolor{blue}{$\downarrow$} \textcolor{blue}{55.24\%} & \textcolor{blue}{$\downarrow$} \textcolor{blue}{59.01\%} & \textcolor{blue}{$\downarrow$} \textcolor{blue}{51.47\%} & \textcolor{blue}{$\downarrow$} \textcolor{blue}{52.38\%} \\\hline
\multicolumn{12}{c}{\underline{\textbf{Rician Channel}}}\\ 
\textbf{Audio (only)}& 0.1624 & 0.1607 & 0.1680 & 0.1687 & 0.1753 & 0.1816 & 0.1801 & 0.1850 & 0.1830 & \underline{\textbf{0.1901}} & 0.1820\\
& &  &  &  &  & \textcolor{blue}{$\downarrow$}\textcolor{blue}{{0.32\%}} & \textcolor{blue}{$\downarrow$}\textcolor{blue}{1.20\%} &  & \textcolor{blue}{$\downarrow$}\textcolor{blue}{3.18\%} &  & \textcolor{blue}{$\downarrow$}\textcolor{blue}{4.31\%} \\
\textbf{Video (only)}& 0.1735 & 0.1810 & 0.1852 & 0.1906 & 0.1919 & 0.1896 & 0.1927 & 0.1977 & 0.2016 & 0.2010 & \underline{\textbf{0.2015}}\\
& & \textcolor{blue}{$\downarrow$}\textcolor{blue}{0.32\%} & \textcolor{blue}{$\downarrow$}\textcolor{blue}{3.38\%} &  & \textcolor{blue}{$\downarrow$}\textcolor{blue}{1.73\%} & \textcolor{blue}{$\downarrow$}\textcolor{blue}{2.28\%} & \textcolor{blue}{$\downarrow$}\textcolor{blue}{3.21\%} & \textcolor{blue}{$\downarrow$}\textcolor{blue}{2.94\%} & \textcolor{blue}{$\downarrow$}\textcolor{blue}{1.02\%} & \textcolor{blue}{$\downarrow$}\textcolor{blue}{0.47\%} & \textcolor{blue}{$\downarrow$}\textcolor{blue}{0.76\%} \\

\textbf{Multimodal}  & 0.1273& 0.2078 & 0.2284& 0.2378& 0.2698& 0.2648 & 0.3063 & \underline{\textbf{0.3758}} & 0.3500 & 0.3557 & 0.3487\\
 & \textcolor{blue}{$\downarrow$} \textcolor{blue}{62.73\%} & \textcolor{blue}{$\downarrow$} \textcolor{blue}{56.86\%} & \textcolor{blue}{$\downarrow$} \textcolor{blue}{55.55\%} & \textcolor{blue}{$\downarrow$} \textcolor{blue}{55.25\%} & \textcolor{blue}{$\downarrow$} \textcolor{blue}{50.50\%} & \textcolor{blue}{$\downarrow$} \textcolor{blue}{53.09\%} & \textcolor{blue}{$\downarrow$} \textcolor{blue}{50.25\%} & \textcolor{blue}{$\downarrow$} \textcolor{blue}{43.38\%} & \textcolor{blue}{$\downarrow$} \textcolor{blue}{49.32\%} & \textcolor{blue}{$\downarrow$} \textcolor{blue}{50.27\%} & \textcolor{blue}{$\downarrow$} \textcolor{blue}{51.47\%} \\
\hline
\end{tabular}
}
\label{tab:SNR_woH}
\end{table*}
\noindent \textbf{Impact of Channel Conditions:} Our method demonstrates substantial resilience to noise interference, maintaining localization accuracy above 0.55 even at low SNR. The performance improves rapidly with increased SNR, showcasing the framework’s ability to handle multipath fading effectively. However, while the method approaches the theoretical upper limit of accuracy, discrepancies remain due to the estimated channel matrix $\mathbf{H}$ not perfectly matching the true $\mathbf{H}$, particularly under high SNR conditions where precise channel state information is crucial.

Overall, our multimodal framework exhibits enhanced effectiveness and robustness, offering superior performance in complex and high-noise communication environments compared to conventional techniques.

\subsubsection{The performance of our multimodal semantic encoder-decoder}
\label{sec:ex_SED}
We begin by illustrating an example of audio-visual event localization in Fig.~\ref{fig:hotmap}. In this example, predicting the event is challenging due to the variability of the visual images and the presence of background noise in the audio signals. 

While both the stable AVE methods (AVEL~\cite{ave} and PSP~\cite{psp}) and our approach utilize AGVA, we demonstrate that our method provides improved attention to visual regions closely related to sound sources. As shown in Fig.~\ref{fig:hotmap}, for the event of male speech, our method focuses on the man, particularly in the initial and final segments. In contrast, AVEL only identifies the background and has very limited receptive fields. Although PSP considers inter-modal enhancement, it overlooks the semantic and temporal variations of audiovisual data in the time-frequency domain. This synchronous mining should be effectively achieved through clear separation and lossless aggregation in the time-frequency domain.

In the second row of the figure, it is evident that while PSP shows better classification results compared to AVEL, it still suffers from the same issue of focusing on the background and having limited receptive fields.

\subsubsection{The performance of our methods without channel matrix estimation}
\label{sec:ex_woCSI}

To validate the effectiveness of the proposed \textit{Complex Module of Pilot-guided Physical Channel and Signal Detection} and address RQ3, we conducted an ablation experiment. This experiment assesses the model's performance without channel matrix estimation using preambles. By maintaining the same semantic and channel encoding at the transmitter (i.e., consistent pre-trained models), we retrained the receiver's model to handle multimodal tasks without the ability to estimate the channel matrix for signal detection and channel-semantic decoding.

As shown in Table~\ref{tab:SNR_woH}, we compared the experimental results of single-modal semantic transmission and multimodal semantic transmission from Fig.~\ref{fig:ex_snr}. The accuracy of each model type under different SNRs clearly indicates a significant performance drop without the channel matrix estimation block, as highlighted by the blue percentage decrease. The multimodal method experiences a larger performance loss in the absence of this module, as it cannot perform channel estimation and can only use noise-distorted and faded signals as input to the receiver. Under these conditions, only the Gaussian channel, which is not affected by fading, shows a relatively smaller performance loss (less than 20\%), while other channels suffer a performance loss exceeding 50\%. Single-modal methods, due to their inherent lack of modality information and the absence of cross-modal enhancement via attention mechanisms, already exhibit lower performance, so the performance loss is less noticeable.

This highlights that existing semantic communication methods, which assume perfect CSI (for Rician fading a fixed state for \(K\)), do not align with real-world conditions where CSI is inherently unknown and needs to be accurately estimated beforehand. Accurate channel state estimation is crucial for effective signal detection, avoiding errors, and demonstrating the significance of the proposed preamble-guided module.

\section{Conclusion}
This paper presents a novel pilot-guided framework for multimodal semantic communication, specifically designed for audio-visual event localization tasks. While multimodal semantic communication enhances efficiency and reliability across fields like AI and smart homes, existing methods often rely on analog channels with constant state assumptions, which are inadequate for dynamic real-world conditions. Our framework addresses these limitations by using digital pilot codes and Euler-based encoding/decoding, adapting to real-time channel states and effectively handling multimodal stream data. Extensive experiments demonstrate that our framework outperforms existing methods in Signal-to-Noise Ratio (SNR), showing superior robustness and performance under variable channel conditions. This framework offers a significant advancement in multimodal semantic communication, providing a robust and versatile solution for real-world applications.
\bibliographystyle{IEEEtran} 
\bibliography{conference_101719} 


\end{document}